\documentclass{iau}
\usepackage{graphicx}
\usepackage{xspace}

\title[Modelling thermal X-rays around the Galactic centre] 
{Modelling the thermal X-ray emission around the Galactic centre from \\colliding Wolf-Rayet winds}

\author[Christopher M.~P.~Russell, Q.~Daniel Wang \& Jorge Cuadra]   
{Christopher M.~P.~Russell$^1$,
 Q.~Daniel Wang$^2$,
 \and Jorge Cuadra$^3$}

\affiliation{$^1$X-ray Astrophysics Laboratory, NASA/Goddard Space Flight Center,\\ Greenbelt, MD 20771, USA (NASA Postdoctoral Program Fellow, administered by USRA)\\ email: {\tt crussell@udel.edu} \\[\affilskip]
$^2$Department of Astronomy, University of Massachusetts, Amherst, MA 01003, USA
\\[\affilskip]
$^3$Instituto de Astrof\'{\i}sica, 
Pontificia Universidad Cat\'{o}lica de Chile, 782-0436 Santiago, Chile
}

\pubyear{2017}
\volume{329}  
\jname{The lives and death-throes of massive stars}
\editors{J.J. Eldridge, L. Xiao, L.A.S. McClelland \\\& J.C. Bray, eds.}

\newcommand\arcsec{\hbox{$^{\prime\prime}$}}   
\newcommand\arcse{\hbox{$^{\prime\prime}$}\xspace}   
\newcommand*{\SAs}{{Sgr\,A$^*$}\xspace}

\begin{document}

\maketitle

\begin{abstract}
  We compute the thermal X-ray emission from hydrodynamic simulations of the 30 Wolf-Rayet (WR) stars orbiting within a parsec of \SAs, with the aim of interpreting the \textit{Chandra} X-ray observations of this region.  The model well reproduces the spectral shape of the observations, indicating that the shocked WR winds are the dominant source of this thermal emission.  The model X-ray flux is tied to the strength of the \SAs outflow, which clears out hot gas from the vicinity of \SAs.  A moderate outflow best fits the present-day observations, even though this supermassive black hole (SMBH) outflow ended $\sim$100\,yr ago.
\end{abstract}

\begin{figure}[h]
\begin{center}
  \includegraphics[height=6.57cm]{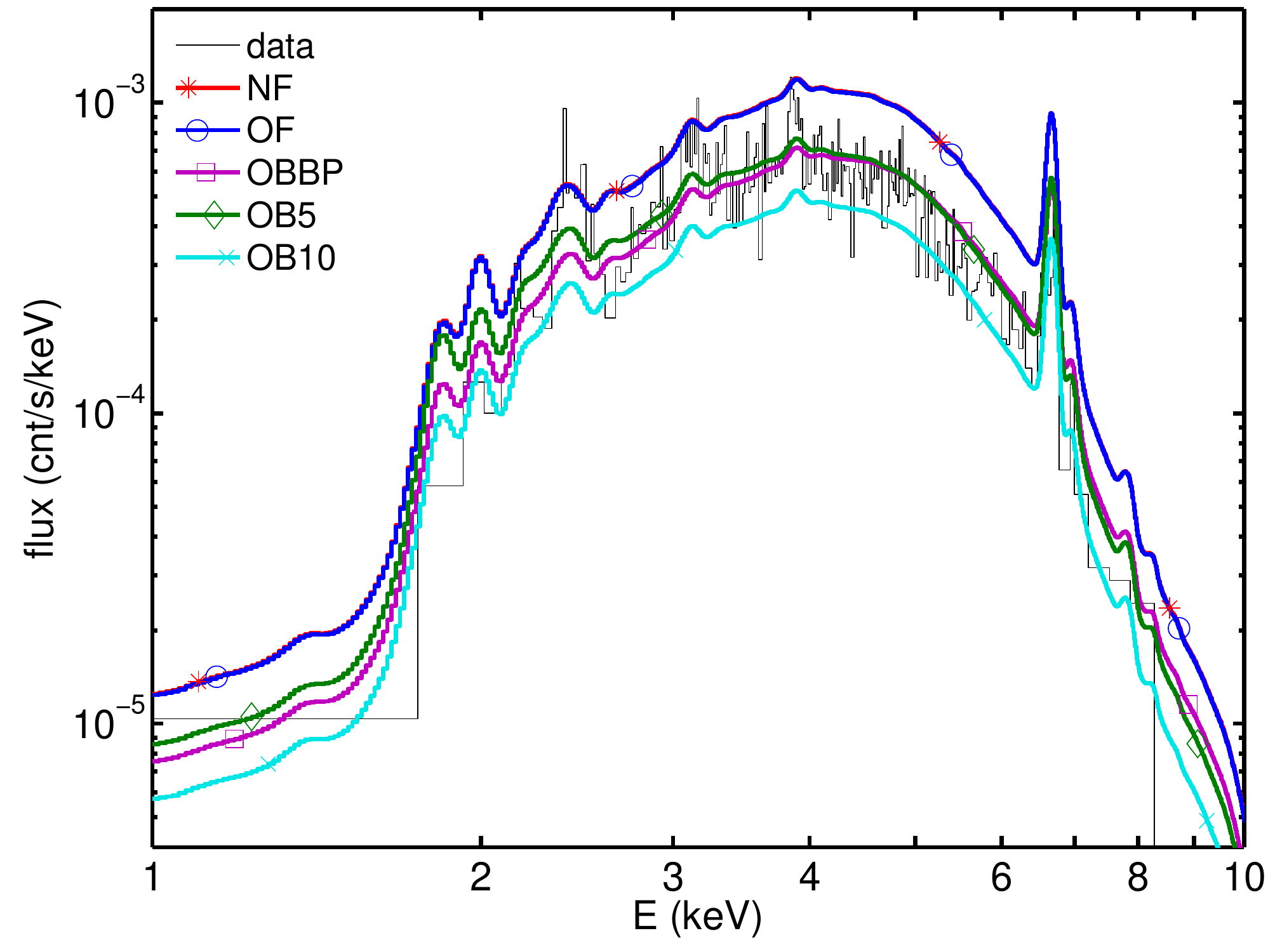}
  \includegraphics[height=6.57cm]{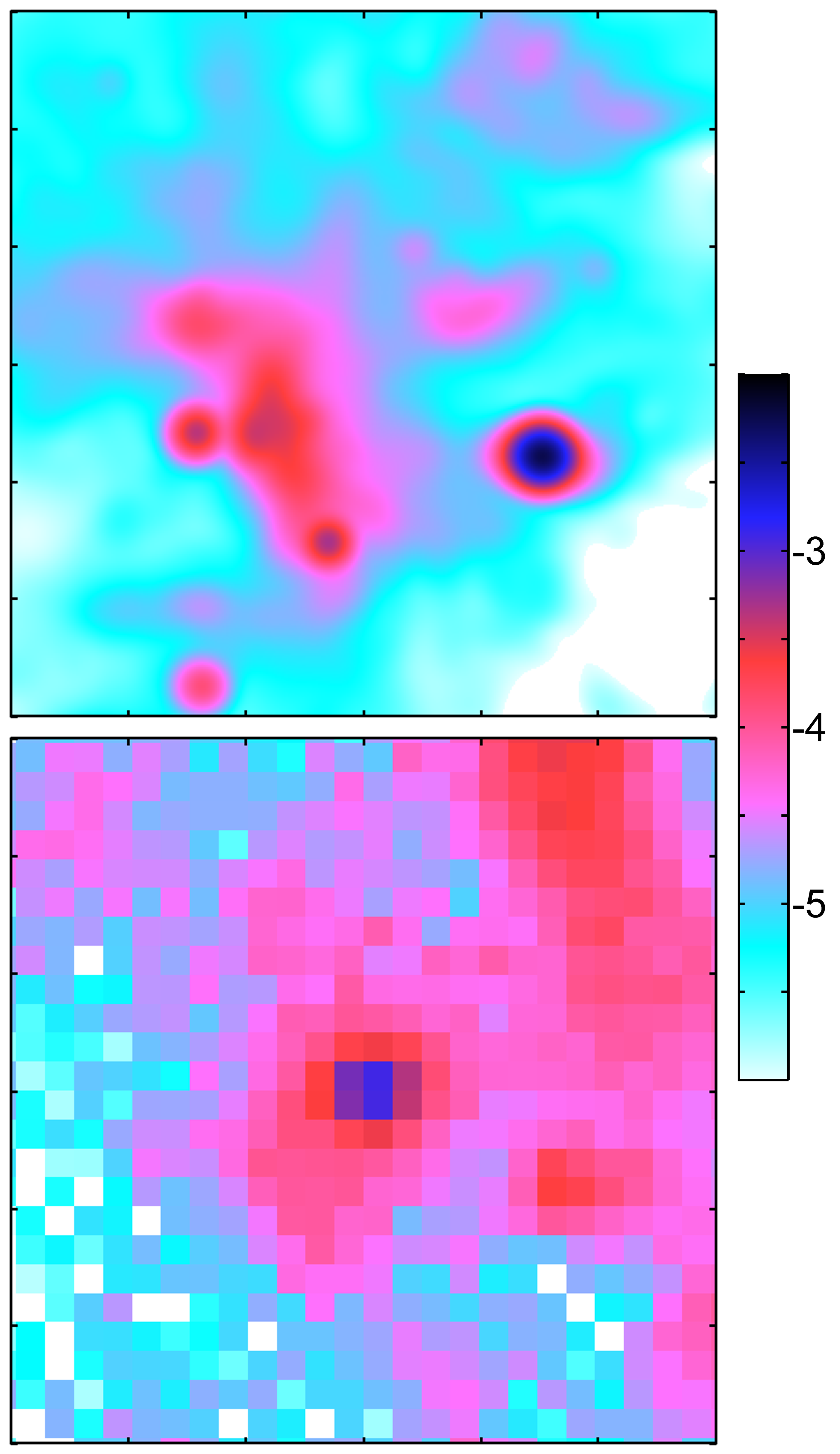}
  \put(-105,106){\scriptsize \fontfamily{phv}\selectfont \textbf{model}}
  \put(-105,98){\scriptsize \fontfamily{phv}\selectfont \textbf{OB5}}
  \put(-105,4){\scriptsize \fontfamily{phv}\selectfont \textbf{data}}
  \put(-38,83){\scriptsize \fontfamily{phv}\selectfont \rotatebox{300}{\textbf{PWN}}}
  \put(3,65){\scriptsize \fontfamily{phv}\selectfont \rotatebox{90}{log cnt/s/arcsec\textsuperscript{\tiny 2}}}

  \caption{\textit{Left}: ACIS-S/HETG zeroth-order spectra from 2\arcse--5\arcse ring around \SAs.  The SMBH outflow increases from NF (no feedback) to OB10 (strongest feedback). \textit{Right}: ACIS-S/HETG zeroth-order 4-9 keV images of 12\arcsec$\times$12\arcse centered on \SAs, showing the best-fit model (OB5) and the data.  The non-thermal emission from the SMBH and pulsar wind nebula is not modeled.}
  \label{fig1}
\end{center}
\end{figure}
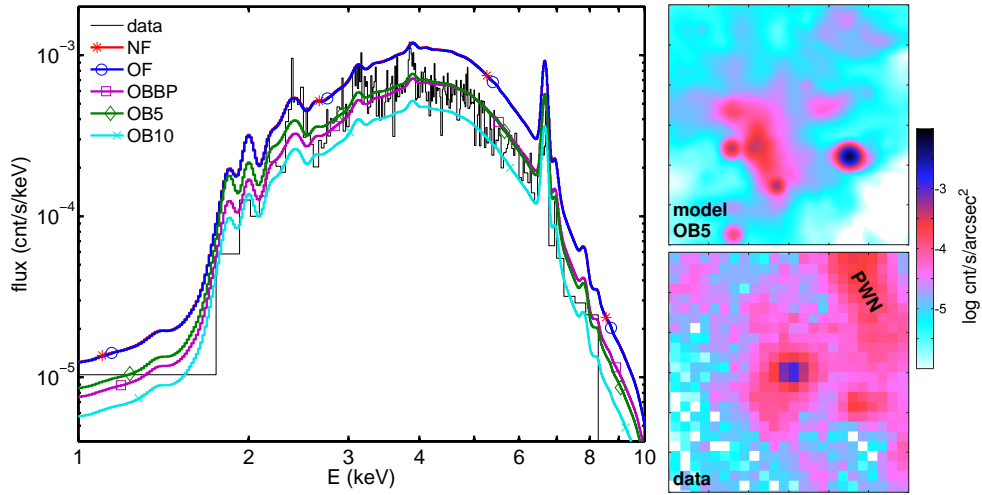

Fig.\,\ref{fig1} shows the main results. See \cite[Russell et al. (2017)]{Russell_etal17} for further details of this work.

\begin{thebibliography}{}

\bibitem[Russell, Wang \& Cuadra (2017)]{Russell_etal17}
{Russell, C.M.P., Wang, Q.D., \& Cuadra, J.} 2017,
\textit{MNRAS}, 464, 4958

\end{thebibliography}
\end{document}